\documentclass[preprint,12pt]{elsarticle}

\usepackage{amssymb}
\usepackage{amsmath}
\usepackage{braket}

\newcommand{\lie}{\mathsterling\!}

\newcommand{\pd}{\partial}

\def\({\left(}
\def\){\right)}
\def\<{\left<}
\def\>{\right>}
\newcommand{\be}{\begin{equation}}
\newcommand{\ee}{\end{equation}}
\newcommand{\bmat}{\left(\begin{matrix}}
\newcommand{\emat}{\end{matrix}\right)}

\journal{Nuclear Physics B}

\begin{document}

\begin{frontmatter}



\title{Higher Derivative Theories for Interacting Massless Gravitons in Minkowski Spacetime}


\author{Dong Bai \corref{cor}}
\ead{dbai@itp.ac.cn}
\address{School of Physics, Nanjing University,  Nanjing, 210093, China}
\cortext[cor]{Corresponding author}

\author{Yu-Hang Xing \corref{corr}}
\ead{xingyh@itp.ac.cn}
\address{Institute of Theoretical Physics, Chinese Academy of Sciences, Beijing 100190, China\\School of Physical Sciences, University of Chinese Academy of Sciences, No.19A Yuquan Road, Beijing 100049, China}

\begin{abstract}
We study a novel class of higher derivative theories for interacting massless gravitons in Minkowski spacetime. These theories were first discussed by Wald decades ago, and are characterized by scattering amplitudes essentially different from general relativity and many of its modifications. We discuss various aspects of these higher derivative theories, including the Lagrangian construction, violation of asymptotic causality, scattering amplitudes, non-renormalization, and possible implications on emergent gravitons from condensed matter systems. 

\end{abstract}

\begin{keyword}


massless graviton \sep higher derivatives \sep asymptotic causality \sep scattering amplitude
\end{keyword}

\end{frontmatter}


\section{Introduction}
\label{Introduction}

Massless gravitons take the central position in modern theoretical physics. The most well-known theory for interacting massless gravitons is Einstein's general relativity (GR), which enjoys many elegant properties, such as the equivalence principle \footnote{Recent studies on bending of light in quantum gravity suggest that certain version of the equivalence principle could be violated at the quantum level \cite{Bjerrum-Bohr:2014zsa,Bjerrum-Bohr:2015vda,Bjerrum-Bohr:2016hpa,Bai:2016ivl,Bjerrum-Bohr:2017dxw}. }, minimal couplings, diffeomorphism invariance, asymptotic causality, simple scattering amplitudes, etc. It was first noted by Wald in the 1980s that not all the theories for interacting massless gravitons consistent with Lorentz invariance and quantum mechanics have to be GR-like \cite{Wald:1986bj,Cutler:1986dv,Wald:1986dw}. There is indeed a second class of higher derivative theories \footnote{In literature, higher derivative theories also refer to higher derivative modifications to GR, which is unfortunate for our studies here. To avoid unnecessary confusions, in this note, higher derivative modifications to GR shall be referred to explicitly by names like $R^2$, $R^3$ modifications to GR instead.} for interacting massless gravitons in Minkowski spacetime characterized by novel scattering amplitudes essentially different from GR and many of its modifications. Additionally, these theories could be defined exactly in Minkowski spacetime without needing to introduce the notion of curved spacetime. Therefore, compared with GR, they look much more like the traditional field theories like Maxwell's theory on electromagnetism and Yang-Mills theory.

In this note, we would like to study properties of these higher derivative theories of interacting massless gravitons. Previous studies on this topic could be found in Ref.~\cite{Wald:1986bj,Cutler:1986dv,Wald:1986dw,Heiderich:1988yq,Heiderich:1990bb,Boulanger:2000rq,Horava:2010zj,deRham:2012ew,Hinterbichler:2013eza}. Recently, Ref.~\cite{Hertzberg:2016djj,Hertzberg:2017abn} give comprehensive treatments on the matter couplings in these higher derivative theories. Compared with previous studies, our work  concentrates on the Lagrangian construction, violation of asymptotic causality, scattering amplitudes, non-renormalization properties. Besides purely academic interests, we would like to suggest that these higher derivative theories might be helpful for studying emergent gravitons in condensed matter systems. This point will be expanded in Section \ref{EmergentGravitons}.

The rest part of this note is organized as follows: In Section \ref{Lagrangians} we study the Lagrangian construction of these higher derivative theories of massless gravitons in Minkowski spacetime. Under reasonable conditions, we show that only two Lagrangian constructions are allowed for the three-point vertices. We show further how these two structures could be generalized to $n$-point vertices. In Section \ref{AsymptoticCausality}, we show that the $S^{(3)}_4$ and $S^{(3)}_6$ theory are incompatible with asymptotic causality. Although we borrow our techniques from Ref.~\cite{Camanho:2014apa}, the situations we encounter here are different from the $R^2$ and $R^3$ modifications to GR studied by the aforementioned reference. The violation of asymptotic causality found here turns out to be more severe. In Section \ref{ScatteringAmplitudes}, we study scattering amplitudes of these higher derivative theories. Our calculations are done in $D=4$ only in order to make use of the spinor-helicity formalism. We calculate explicitly the four-point scattering amplitudes of the $S^{(3)}_6$ theory, as well as their large $z$ behaviors under Britto-Cachazo-Feng-Witten (BCFW) shift \cite{Britto:2004ap,Britto:2005fq}. We identify a specific helicity configuration which receives all its contributions from the boundary of the complex $z$-space. Although the calculations are done for the $S^{(3)}_6$ theory, the results could be helpful for studies of $R^3$ modifications of GR as well. In Section \ref{NonRenormalization}, we study the non-renormalization property of these higher derivative theories. It is shown explicitly that the $S^{(3)}_4$ and $S^{(3)}_6$ vertices are immune to quantum corrections. In Section \ref{EmergentGravitons}, we remark on the possible implications of these higher derivative theories on emergent gravitons from condensed matter systems. We end this note with further directions in Section \ref{Summary}. In \ref{PseudoLinear}, we provide a proof on the linearized diffeomorphism invariance of the so-called pseudo-linear terms.

\section{Lagrangians} 
\label{Lagrangians}

In this section, we would like to construct Lagrangians for the novel higher derivative theories of interacting massless gravitons. Explicitly, we are looking for Lagrangians respecting the linearized diffeomorphism invariance, which is a gauge symmetry. Although widely regarded as theoretical redundancy, gauge invariance could still be helpful in the Lagrangian approach to quantum field theories. This point has been verified by the historical developments of gauge theories, and here we shall continue to utilize this traditional wisdom. 

The kinematic terms for massless gravitons in dimensions $D\ge4$ are given by 
\begin{align}
S^{(2)}&=\int\mathrm{d}^Dx \left(\sqrt{-g}R\right)^{(2)}\nonumber\\
&=\int\mathrm{d}^Dx \left(\frac{1}{2}h^{\mu\nu}\partial_{\rho}\partial^{\rho}h_{\mu\nu} - \frac{1}{2}h^{\mu}_{\mu}\partial_{\rho}\partial^{\rho}h^{\nu}_{\nu} - h^{\mu\nu}\partial_{\rho}\partial_{\nu}h_{\mu}^{\rho} + h^{\mu}_{\mu}\partial_{\rho}\partial_{\nu}h^{\nu\rho}\right),
\label{FierzPauli}
\end{align}
which is well-known to be invariant under linearized diffeomorphism. $\left(\sqrt{-g}R\right)^{(2)}$ in the first line is used only as a shorthand, and this does not mean that our discussions have anything indispensable to do with the curved spacetime.

Since $S^{(2)}$ describes free gravitons only, extra vertices are needed in order to have interacting theories. In this note, we shall mainly concentrate on three-point vertices. It is pointed out by Ref.~\cite{Camanho:2014apa} that in dimensions $D\ge4$, there are only three types of parity-preserving \emph{on-shell} three-point amplitudes of massless gravitons:
\begin{align}
&\mathcal{A}^{(3)}_2 = (\epsilon_1.\epsilon_2\,\epsilon_3.p_1 + \epsilon_1.\epsilon_3\,\epsilon_2.p_3 + \epsilon_2.\epsilon_3\,\epsilon_1.p_2)^2,\nonumber\\
&\mathcal{A}^{(3)}_4 = (\epsilon_1.\epsilon_2\,\epsilon_3.p_1 + \epsilon_1.\epsilon_3\,\epsilon_2.p_3 + \epsilon_2.\epsilon_3\,\epsilon_1.p_2)\epsilon_1.p_2\,\epsilon_2.p_3\,\epsilon_3.p_1,\nonumber\\
&\mathcal{A}^{(3)}_6 = (\epsilon_1.p_2\,\epsilon_2.p_3\,\epsilon_3.p_1)^2.\nonumber
\end{align}
Here, we have used the replacing rule $\epsilon_{\mu\nu}\to\epsilon_{\mu}\epsilon_{\nu}$ for massless gravitons. The subscripts ``2'', ``4'', ``6'' denote the numbers of momenta in these three amplitudes. Noticeably, $\mathcal{A}^{(3)}_2$ corresponds to the three-point amplitude in GR. The most general three-point amplitude is given by a linear combination of the above three amplitudes
\begin{equation}
\mathcal{A}^{(3)}_{hhh} = \alpha_2 \mathcal{A}^{(3)}_2 + \alpha_4 \mathcal{A}^{(3)}_4 + \alpha_6 \mathcal{A}^{(3)}_6.\nonumber
\end{equation} 
Usually, $\alpha_2$ is taken to be non-vanishing, which makes the resulting theories be inevitably GR-like. It is interesting to ask \emph{whether this is really inevitable}. The answer turns out to be NO, and is directly related to higher derivative theories under construction. 

Some helpful results could be summarized in the following:
\begin{itemize}
\item There is no dimensional independent three-point vertex giving rise to the on-shell three-point amplitude $\mathcal{A}^{(3)}_2$, while preserving linearized diffeomorphism. In other words, the appearance of $\mathcal{A}^{(3)}_2$ is the smoking gun for the gravitational theory to be GR-like.
\item There is only one dimensional independent three-point vertices giving rise to the on-shell three-point amplitude $\mathcal{A}^{(3)}_4$,
\begin{equation}
S^{(3)}_4 = \int\mathrm{d}^Dx\ \delta^{\mu_1\nu_1\mu_2\nu_2 \alpha}_{\rho_1\sigma_1\rho_2\sigma_2 \beta}\ h_\alpha^\beta\pd_{\mu_1}\pd^{\rho_1}h_{\nu_1}^{\sigma_1}\pd_{\mu_2}\pd^{\rho_2}h_{\nu_2}^{\sigma_2},
\label{H3P4}
\end{equation}
\noindent which is exactly the cubic order expansion in $h_{\mu\nu}$ of the Gauss-Bonnet term $\sim \sqrt{-g}(R^2 - 4R^{\mu\nu}R_{\mu\nu} + R^{\mu\nu\rho\sigma}R_{\mu\nu\rho\sigma})$. This is also a special case of the so-called pseudo-linear terms \cite{Hinterbichler:2013eza}. A formal demonstration of why such structures respect linearized diffeomorphism is given in \ref{PseudoLinear}.
\item After choosing the proper basis, the only dimensional independent three-point vertices giving rise to the on-shell three-point amplitude $\mathcal{A}^{(3)}_6$ is given by
\begin{equation}
S^{(3)}_6 = \int\mathrm{d}^Dx R^{(1)\mu\nu}_{\ \ \ \ \ \alpha\beta}R^{(1)\alpha\beta}_{\ \ \ \ \ \rho\sigma}R^{(1)\rho\sigma}_{\ \ \ \ \ \mu\nu},
\label{H3P6}
\end{equation}
where $R^{(1)}_{\mu\nu\rho\sigma}\equiv -\frac{1}{2}\partial_{\rho}\partial_{\mu}h_{\nu\sigma} + \frac{1}{2}\partial_{\rho}\partial_{\nu}h_{\mu\sigma} + \frac{1}{2}\partial_{\sigma}\partial_{\mu}h_{\nu\rho}-\frac{1}{2}\partial_{\sigma}\partial_{\nu}h_{\mu\rho}$ is the linearized Riemann tensor, and indices are raised up by $\eta^{\mu\nu}$ rather than $g^{\mu\nu}$. 
\end{itemize}
The above results are obtained in a brute-force way with Mathematica. Take the first result as an example. There, we start with linear combinations of all possible scalar contractions with three $h_{\mu\nu}$s and two derivatives $\partial_{\mu}$s, multiplying by various free coefficients. The resulting expression is required to be invariant under linearized diffeomorphism up to total derivatives, which imposes several conditions upon the free coefficients. For three $h_{\mu\nu}$s and two $\partial_{\mu}$s, no consistent choice for these free coefficients could be found, which leads to the first result mentioned above. The second and third result could be obtained similarly by repeating the analysis for three $h_{\mu\nu}$s and four $\partial_{\mu}$s and three $h_{\mu\nu}$s and six $\partial_{\mu}$s, respectively. Although these three properties have been mentioned in some way in, e.g., Ref.~\cite{Metsaev:2005ar}, our derivation here is based on brute-force symbolic computations and can be viewed as a cross-check.

Three technical remarks are given as follows:

1. When constructing and simplifying various tensorial expressions, we haven't taken into considerations any dimensional dependent quantities such as those associated with Levi-Civita symbol $\varepsilon_{\mu_1\cdots\mu_D}$. This is exactly what we would like to convey by ``dimensional independent" in stating the above results.

2. The above results are characterized by both ``existence'' and ``uniqueness'' of linearized diffeomorphism invariant three-point vertices. In the second and third result, some curved-spacetime notations are used to save space. Although the existence of proper three-point vertices could indeed be understood from the curved-spacetime viewpoints, such an understanding is not indispensable as shown by our brute-force treatments. It is, on the contrary, somehow surprising that this is possible to have the curved-spacetime interpretation. Moreover, it is not clear to us yet how the uniqueness could be derived from the curved-spacetime viewpoints.

3. For three $h_{\mu\nu}$s and four $\partial_{\mu}$s, the only linearized diffeomorphism invariant three-point vertices are given by Eq.~\eqref{H3P4}, which give rise to $\mathcal{A}^{(3)}_4$ exactly. On the other hand, for three $h_{\mu\nu}$s and six $\partial_{\mu}$s, we have various choices for the three-point vertices. In fact, as $R^{(1)}_{\mu\nu\rho\sigma}$ and all its contractions turn out to be \emph{exactly} linearized diffeomorphism invariant,  so are all of their cubic scalar contractions. However, not all these three-point vertices give rise to $\mathcal{A}^{(3)}_6$. For instance, $\left(R^{(1)}\right)^3$ leads to vanishing on-shell three-point amplitudes. The three-point vertices giving rise to $\mathcal{A}^{(3)}_6$ are scalar cubic contractions solely made of $R^{(1)}_{\mu\nu\rho\sigma}$s, among which only two are linearly independent. A convenient basis could be $I_1\equiv R^{(1)\mu\nu}_{\ \ \ \ \ \alpha\beta}R^{(1)\alpha\beta}_{\ \ \ \ \ \rho\sigma}R^{(1)\rho\sigma}_{\ \ \ \ \ \mu\nu}$ and $G_3\equiv I_1 - 2 R^{(1)\ \,\alpha\ \beta}_{\ \ \ \mu\ \rho}R^{(1)\mu\nu\rho\sigma}R^{(1)}_{\nu\alpha\sigma\beta}$. It is straightforward to show that the on-shell three-point amplitude given by $G_3$ actually vanishes, which agrees with Ref.~\cite{Metsaev:1986yb}. Thus, it is in this basis that only $I_1$ gives rise to $\mathcal{A}^{(3)}_6$, as shown in the third result.

The brute-force treatments could be ``easily'' generalized to $n$-point vertices, at the expense of inflating computation durations. Examples of $n$-point vertices involve the $n$-point expansion of $(n-1)$-th Lovelock terms \footnote{The $n$-point expansion from Lovelock terms are invariant up to total derivatives under linearized diffeomorphism. $n=2$ corresponds to the kinematic terms $S^{(2)}$, and $n=3$ corresponds to the three-point vertices $S^{(3)}_4$. The characteristic property of these vertices is that they give rise to only second-order equations of motion, and thus are ghost-free.} (see \ref{PseudoLinear}), and scalar contractions of $n$ copies of $R^{(1)}_{\mu\nu\rho\sigma}$, $R^{(1)}_{\mu\nu}$ or $R^{(1)}$. It is not clear yet whether there are extra contributions to $n$-point vertices, and it is interesting to work out some proofs for uniqueness \cite{Bai:2017dwf}. In fact, if additionally we require non-existence of Ostrogradski ghosts, a proof could be done similar to the famous proof by Lovelock for the Lovelock gravity \cite{Lovelock:1971yv,Lovelock:1972vz}.

Up to now, we consider only self-interactions of massless gravity in higher derivative theories. One way to couple massless gravitons to matter is to construct scalar contractions with $R^{(1)}_{\mu\nu\rho\sigma}$s, various matter fields, and spacetime derivatives $\partial_{\mu}$s. For example, the coupling of massless gravitons to photons could be introduced by $R^{(1)}_{\mu\nu\rho\sigma}F^{\mu\nu}F^{\rho\sigma}$. More comprehensive discussions on the matter couplings could be found in Ref.~\cite{Hertzberg:2016djj,Hertzberg:2017abn}. See, also, Ref.~\cite{BeltranJimenez:2018tfy} for a relevant discussion on scalar and vector fields coupled to the energy-momentum tensor.

\section{Asymptotic Causality}
\label{AsymptoticCausality}
Loosely speaking, principle of asymptotic causality states that \emph{interactions always slow you down}. Applications of this principle range from the fact that light travels faster in vacuum than in glass to the celebrating Shapiro time delay \cite{Shapiro:1964uw}. See also Ref.~\cite{Gao:2000ga} for a rigorous treatment of asymptotic causality in GR. Recently, asymptotic causality has been used to study $R^2$ and $R^3$ modifications to GR \cite{Camanho:2014apa}. In this section we would like to use the same techniques to study the $S^{(3)}_4$ and $S^{(3)}_6$ theory \footnote{By the $S^{(3)}_4$ theory, we mean the higher derivative theory whose non-vanishing vertices are given by $S^{(3)}_4$. Similar conventions hold for the $S^{(3)}_6$ theory as well.}.

Asymptotic causality of a target theory could be captured by
\begin{equation}
\delta(\vec{b},s)\equiv\frac{1}{2s}\int\frac{\mathrm{d}^{D-2}\vec{q}}{(2\pi)^{D-2}}e^{i\vec{q}.\vec{b}}\mathcal{A}_4(\vec{q}),
\end{equation} 
where $\mathcal{A}_4(\vec{q})$ is a shorthand for the tree-level four-point amplitude evaluated in momentum and polarization-vector configurations given by: \footnote{In this section, following Ref.~\cite{Camanho:2014apa} we adopt the light-cone coordinate $ds^2 = -dudv + \sum\limits_{i=1}^{D-2}(dx_i)^2$. $\vec{q}$ denotes transverse momentum, $\vec{e}_i$ denotes transverse polarizations, and $\vec{b}$ denotes transverse displacement.}
\begin{align}
&p_{1\mu} = \left(p_u,\,\frac{q^2}{16p_u},\,\frac{\vec{q}}{2}\right),\qquad\qquad\quad\ \ p_{2\mu} = \left(\frac{q^2}{16p_v},\,p_v,\,-\frac{\vec{q}}{2}\right),\nonumber\\
&p_{3\mu} = - \left(p_u,\,\frac{q^2}{16p_u},\,-\frac{\vec{q}}{2}\right),\qquad\qquad p_{4\mu} = - \left(\frac{q^2}{16p_v},\,p_v,\,\frac{\vec{q}}{2}\right),\nonumber\\
&\epsilon^{1\mu} = \left(-\frac{\vec{q}.\vec{e}_1}{2p_u},\,0,\,\vec{e}_1\right),\qquad\qquad\quad\ \ \epsilon^{3\mu} = \left(\frac{\vec{q}.\vec{e}_3}{2p_u},\,0,\,\vec{e}_3\right),\nonumber\\
&\epsilon^{2\mu} = \left(0,\,\frac{\vec{q}.\vec{e}_2}{2p_v},\,\vec{e}_2\right),\qquad\qquad\qquad\ \epsilon^{4\mu} = \left(0,\,-\frac{\vec{q}.\vec{e}_4}{2p_v},\,\vec{e}_4\right),\nonumber
\end{align}
with $s\approx4p_up_v$ and $t\approx-(\vec{q})^2$. The external momenta are chosen in such a way that $s$ is much larger than $t$, but small enough such that the target theory is still weakly coupled. The polarization tensors for massless gravitons can be obtained by products of polarization vectors $\epsilon_{\mu\nu} = \epsilon_{\mu}\epsilon_{\nu}$. Principle of asymptotic causality then requires that $\delta(\vec{b},s)\ge0$ for \emph{all} possible $\vec{b}$ and $\vec{e}_i$ choices.

Using on-shell methods \cite{Camanho:2014apa}, $\delta(\vec{b},s)$ could be calculated as follows:
\begin{equation}
\delta(\vec{b},s) =
\begin{cases}
&\!\!\!\!\!\frac{\Gamma(\frac{D-4}{2})}{4\pi^{\frac{D-2}{2}}}\sum\limits_I\frac{\mathcal{A}^{(3)}_{13I}(-i\partial_{\vec{b}})\mathcal{A}^{(3)}_{I24}(-i\partial_{\vec{b}})}{2s}\frac{1}{|\vec{b}|^{D-4}},\quad(D>4)\\
&\\
&\!\!\!\!\! \frac{1}{2 \pi}\sum\limits_I\frac{\mathcal{A}^{(3)}_{13I}(-i\partial_{\vec{b}})\mathcal{A}^{(3)}_{I24}(-i\partial_{\vec{b}})}{2s}(-\log{|\vec{b}|}),\quad(D=4)
\end{cases}
\label{DeltaD}
\end{equation}
where $\mathcal{A}^{(3)}_{13I}(-i\partial_{\vec{b}})$ and $\mathcal{A}^{(3)}_{I24}(-i\partial_{\vec{b}})$ are related to on-shell three-point amplitudes, whose expressions are model-dependent. For $S^{(3)}_4$ and $S^{(3)}_6$, we have
\begin{align}
&S^{(3)}_4:\quad\sum\limits_I\mathcal{A}^{(3)}_{13I}(-i\partial_{\vec{b}})\mathcal{A}^{(3)}_{I24}(-i\partial_{\vec{b}}) = s^2 e_1^{ij}e_3^{ik}e_2^{lm}e_4^{ln}\partial_{b^j}\partial_{b^k}\partial_{b^m}\partial_{b^n},
 \label{R2Operator}\\
&S^{(3)}_6:\quad\sum\limits_I\mathcal{A}^{(3)}_{13I}(-i\partial_{\vec{b}})\mathcal{A}^{(3)}_{I24}(-i\partial_{\vec{b}}) = s^2 e_1^{ij}e_3^{kl}e_2^{mn}e_4^{pq}\partial_{b^i}\partial_{b^j}\partial_{b^k}\partial_{b^l}\partial_{b^m}\partial_{b^n}\partial_{b^p}\partial_{b^q},
\label{R3Operator}
\end{align}
respectively.

In the rest parts of this section, we would like to show that the $S^{(3)}_4$ and $S^{(4)}_6$ theory violate asymptotic causality, in much the same way that the $R^2$ and $R^3$ modification to GR would violate asymptotic causality as shown in Ref.~\cite{Camanho:2014apa}. To see asymptotic causality violation for the $S^{(3)}_4$ theory, we could take the following choice of $\vec{b}$ and $\vec{e}_i$:
\begin{equation}
\vec{e}_3=\vec{e}_1,\quad \vec{e}_4=\vec{e}_2,\quad e_{1xy} = e_{1yx} = 1,\quad e_{2yz} = e_{2zy} = 1,\quad \vec{b}=(b,\,0,\cdots,\,0).
\label{R2Parameter}
\end{equation}
Take \eqref{R2Operator} and \eqref{R2Parameter} back to Eq.~\eqref{DeltaD}, we have 
\begin{equation*}
\delta_4(\vec{b},s)=  
\begin{cases}
&\!\!\!\!\!- \frac{\Gamma(\frac{D-4}{2})s}{4\pi^{\frac{D-2}{2}}b^D} (D-4) (D-3) (D-2) <0,\quad(D>4)\\
&\\
&\!\!\!\!\! -\frac{s}{\pi b^4}<0,\quad(D=4)
\end{cases}
\end{equation*}
which contradicts the requirement of asymptotic causality $\delta(\vec{b},s)\ge0$.

Similarly, for the $S^{(3)}_6$ theory we could take the following choice of $\vec{b}$ and $\vec{e}_i$:
\begin{equation}
\vec{e}_3=\vec{e}_1,\quad \vec{e}_4=\vec{e}_2,\quad e_{1xx} = - e_{1yy} = 1,\quad e_{2xy} = e_{2yx} = 1,\quad \vec{b}=(b,\,0,\cdots,\,0).
\label{R3Parameter}
\end{equation}
Take \eqref{R3Operator} and \eqref{R3Parameter} back to Eq.~\eqref{DeltaD}, we have 
\begin{equation*}
\delta_6(\vec{b},s) = 
\begin{cases}
&\!\!\!\!\!- \frac{\Gamma(\frac{D-4}{2})s}{2\pi^{\frac{D-2}{2}}b^{D+4}}(D-4)(D-2)D(D+2)(D+3)(D^2+6D+20)<0,\ (D>4)\\
&\\
&\!\!\!\!\!- \frac{s}{4\pi}\frac{80640}{b^8}<0,\quad(D=4)
\end{cases}
\end{equation*}
which contradicts the requirement of asymptotic causality $\delta(\vec{b},s)\ge0$.

Several remarks are given as follows:

1. Higher derivative theories of massless gravitons with non-vanishing on-shell three-point amplitudes $\mathcal{A}^{(3)}_4$ and $\mathcal{A}^{(3)}_6$ turn out to be \emph{inconsistent} with principle of asymptotic causality. It is interesting to note that the causality violation encountered here is even more severe than that in the $R^2$ and $R^3$ modifications to GR. In the later case, the GR's three-point vertex respects asymptotic causality by itself, and the causality violations from $R^2$ and $R^3$ terms become a problem only when they are comparable to the GR's term. This could be achieved only when the transverse distance $b$ is small enough. However, in our case the screening of GR no longer exist, and the causality violation comes into being even when $b$ is large. Asymptotic causality singles out GR and many of its modifications as the only legitimate theories for pure gravity theories with non-vanishing on-shell three-point amplitudes. This result could be viewed as a valuable complement to various ``proofs'' of the uniqueness of GR \cite{Deser:1969wk,Deser:1987uk,Padmanabhan:2004xk,Deser:2009fq}.

2. One naive approach to avoid constraints from asymptotic causality could be requiring $S^{(3)}_4$ or $S^{(3)}_6$ vertices to be vanishing. Even in this case, massless gravitons could interact with each other through either higher-point vertices, or three-point vertices that do not contribute to $\mathcal{A}^{(3)}_4$ or $\mathcal{A}^{(3)}_6$, say, $\left(R^{(1)}\right)^3$, which makes only \emph{off-shell} contributions to higher-point amplitudes.

3. A sophisticated solution could be adding an infinite number of massive higher spin ($> 2$) particles, as argued by Ref.~\cite{Camanho:2014apa}. In this case, the $S^{(3)}_4$ and $S^{(3)}_6$ theory could be used to describe the massless sector of the full theory.

4. A radical possibility could be that these higher derivative theories might provide an effective description for emergent gravitons from condensed matter systems. If this is the case, the violation of asymptotic causality could simply be a reflection of the lack of the notion of relativistic causality in underlying non-relativistic condensed matter systems.

\section{Scattering Amplitudes} 
\label{ScatteringAmplitudes}

Scattering amplitudes encode lots of informations of interactions of massless particles, sometimes even putting constraints on the corresponding UV completions \cite{Bai:2016pae}. In this section, we shall study scattering amplitudes in higher derivative theories, which are largely ignored by previous studies. We shall consider scattering amplitudes in $D=4$ only, as the results could be largely simplified by using spinor-helicity formalism. 

For the $S^{(3)}_4$ theory, the three-point amplitude $\mathcal{A}^{(3)}_4$ could be written as
\begin{equation}
\mathcal{A}^{(3)}_4 = \mathcal{A}^{(3)}_{\text{YM}} \times \mathcal{A}^{(3)}_{\text{SS}}.\nonumber
\end{equation}
Here, $\mathcal{A}^{(3)}_{\text{YM}}$ and $\mathcal{A}^{(3)}_{\text{SS}}$ are on-shell three-point amplitudes of Yang-Mills and Scherk-Schwarz theory respectively. By Scherk-Schwarz theory, we refer to the Lagrangian given by
\begin{equation}
S_{SS} = \int\mathrm{d}^Dx \left[-\frac{1}{4}\text{Tr}(F_{\mu}^{\nu}F^{\mu}_{\nu}) + \text{Tr}(F_{\mu}^{\nu}F_{\nu}^{\rho}F_{\rho}^{\mu})\right],
\label{ScherkSchwarz}
\end{equation}
with $F_{\mu\nu}\equiv F^a_{\mu\nu}T^a$ as the field strength, and $F^a_{\mu\nu}\equiv \partial_{\mu}A^{a}_{\nu}-\partial_{\nu}A^{a}_{\mu}$. For simplicity, the coupling constant is set to be 1. Physically, Scherk-Schwarz theory describes a collection of massless photons (rather than gluons). As far as we know, this three-point vertex $\text{Tr}(F_{\mu}^{\nu}F_{\nu}^{\rho}F_{\rho}^{\mu})$ first appeared in discussions of bosonic-string modifications of Yang-Mills theory made by Scherk and Schwarz \cite{Scherk:1974ca}. Eq.~\eqref{ScherkSchwarz} is named after them to commemorate this achievement. Then, it is straightforward to show that $\mathcal{A}^{(3)}_4=0$ in $D=4$, as $\mathcal{A}^{(3)}_{\text{YM}}$ is non-vanishing only for $(-,-,+)$ and $(+,+,-)$, while $\mathcal{A}^{(3)}_{\text{SS}}$ is non-vanishing only for $(-,-,-)$ and $(+,+,+)$. Higher-point amplitudes could be calculated by Feynman diagrams, and it turns out that they vanish for all possible helicity configurations. This, in fact, reproduces the well-known fact that $\left[\sqrt{-g}(R^2 - 4R^{\mu\nu}R_{\mu\nu} + R^{\mu\nu\rho\sigma}R_{\mu\nu\rho\sigma})\right]^{(3)}$ is a total derivative in $D=4$ and cannot give rise to non-vanishing matrix elements.

On the other hand, $S^{(3)}_6$ could indeed give rise to nontrivial scattering amplitudes in $D=4$. Similarly, for three-point amplitude one has
\begin{equation}
\mathcal{A}^{(3)}_6 = \mathcal{A}^{(3)}_{\text{SS}}\times\mathcal{A}^{(3)}_{\text{SS}}.\nonumber 
\end{equation}
which is non-vanishing for $(-,-,-)$ and $(+,+,+)$ only. Explicitly, we have
\begin{equation}
\mathcal{A}^{(3)}_6(1^-,2^-,3^-) = \left(\braket{12}\braket{23}\braket{31}\right)^2,\quad\mathcal{A}^{(3)}_6(1^+,2^+,3^+) = \left([12][23][31]\right)^2\nonumber
\end{equation}
The four-point amplitudes could be obtained by Feynman diagrammatic calculations, and non-vanishing amplitudes (up to parity conjugations and cyclic permutations) are listed as follows:
\begin{align}
&\mathcal{A}^{(4)}_6(1^-,2^-,3^+,4^+) =\frac{\braket{12}\braket{13}\braket{14}\braket{23}\braket{24}[43]^6}{[12]},\nonumber\\
&\mathcal{A}^{(4)}_6(1^-,2^+,3^-,4^+) = \frac{\braket{12}\braket{13}\braket{14}\braket{23}\braket{34}[42]^6}{[31]},\nonumber\\
&\mathcal{A}^{(4)}_6(1^-,2^-,3^-,4^-) = -\frac{2}{s_{13}^3}\Bigg\{\frac{s_{12}(s_{12}^4+s_{12}^3s_{13}-3s_{12}^2s_{13}^2-7s_{12}s_{13}^3-3s_{13}^4)}{s_{12}+s_{13}}\braket{14}^4\braket{23}^4\nonumber\\
&+(4s_{12}^4+s_{12}^3s_{13}-10s_{12}^2s_{13}^2-19s_{12}s_{13}^3-8s_{13}^4)\braket{12}\braket{14}^3\braket{23}^3\braket{34}\nonumber\\
&+\frac{(s_{12}+s_{13})(6s_{12}^4-3s_{12}^3s_{13}-10s_{12}^2s_{13}^2-16s_{12}s_{13}^3-s_{13}^4)}{s_{12}}\braket{12}^2\braket{14}^2\braket{23}^2\braket{34}^2\nonumber\\
&+\frac{4s_{12}^6+3s_{12}^5s_{13}-8s_{12}^4s_{13}^2-19s_{12}^3s_{13}^3-14s_{12}^2s_{13}^4+s_{13}^6}{s_{12}^2}\braket{12}^3\braket{14}\braket{23}\braket{34}^3\nonumber\\
&+\frac{(s_{12}+s_{13})(s_{12}^6-2s_{12}^4s_{13}^2-6s_{12}^3s_{13}^3-3s_{12}^2s_{13}^4+s_{13}^6)}{s_{12}^3}\braket{12}^4\braket{34}^4\nonumber
\Bigg\}.\nonumber
\end{align}

The large $z$ behavior of four-point amplitudes under BCFW shift \cite{Britto:2004ap,Britto:2005fq}
\begin{equation}
\ket{i}\to\ket{\hat{i}}\equiv\ket{i}-z \ket{j},\qquad |j]\to|\check{j}]\equiv|j]+z |i]
\end{equation} 
is then given by
\begin{align}
\!\!\!\!\!\!\!\!\!\!\!\!\!&\mathcal{A}^{(4)}_6(\hat{1}^-,2^-,\check{3}^+,4^+),\, \mathcal{A}^{(4)}_6(\hat{1}^-,2^-,3^+,\check{4}^+),\, \mathcal{A}^{(4)}_6(1^-,\hat{2}^-,\check{3}^+,4^+),\, \mathcal{A}^{(4)}_6(1^-,\hat{2}^-,3^+,\check{4}^+) \sim z^8,\nonumber\\
\!\!\!\!\!\!\!\!\!\!\!\!\!&\mathcal{A}^{(4)}_6(\hat{1}^-,\check{2}^-,3^+,4^+),\, \mathcal{A}^{(4)}_6(\check{1}^-,\, \hat{2}^-,3^+,4^+),\, \mathcal{A}^{(4)}_6(1^-,2^-,\hat{3}^+,\check{4}^+),\, \mathcal{A}^{(4)}_6(1^-,2^-,\check{3}^+,\hat{4}^+) \sim z^2,\nonumber\\
\!\!\!\!\!\!\!\!\!\!\!\!\!&\mathcal{A}^{(4)}_6(\check{1}^-,2^-,\hat{3}^+,4^+),\, \mathcal{A}^{(4)}_6(1^-,\check{2}^-,\hat{3}^+,4^+),\, \mathcal{A}^{(4)}_6(\check{1}^-,2^-,3^+,\hat{4}^+),\, \mathcal{A}^{(4)}_6(1^-,\check{2}^-,3^+,\hat{4}^+) \sim z^0;\nonumber\\
\!\!\!\!\!\!\!\!\!\!\!\!\!&\mathcal{A}^{(4)}_6(\hat{1}^-,\check{2}^+,3^-,4^+),\, \mathcal{A}^{(4)}_6(\hat{1}^-,2^+,3^-,\check{4}^+),\,\mathcal{A}^{(4)}_6(1^-,\check{2}^+,\hat{3}^-,4^+),\, \mathcal{A}^{(4)}_6(1^-,2^+,\hat{3}^-,\check{4}^+) \sim z^8,\nonumber\\
\!\!\!\!\!\!\!\!\!\!\!\!\!&\mathcal{A}^{(4)}_6(\hat{1}^-,2^+,\check{3}^-,4^+),\, \mathcal{A}^{(4)}_6(1^-,\hat{2}^+,3^-,\check{4}^+),\, \mathcal{A}^{(4)}_6(\check{1}^-,2^+,\hat{3}^-,4^+),\,\mathcal{A}^{(4)}_6(1^-,\check{2}^+,3^-,\hat{4}^+) \sim z^2,\nonumber\\
\!\!\!\!\!\!\!\!\!\!\!\!\!&\mathcal{A}^{(4)}_6(\check{1}^-,\hat{2}^+,3^-,4^+),\, \mathcal{A}^{(4)}_6(1^-,\hat{2}^+,\check{3}^-,4^+),\,\mathcal{A}^{(4)}_6(\check{1}^-,2^+,3^-,\hat{4}^+),\,\mathcal{A}^{(4)}_6(1^-,2^+,\check{3}^-,\hat{4}^+) \sim z^0;\nonumber\\
\!\!\!\!\!\!\!\!\!\!\!\!\!&\mathcal{A}^{(4)}_6(\hat{1}^-,\check{2}^-,3^-,4^-),\,\mathcal{A}^{(4)}_6(\hat{1}^-,2^-,\check{3}^-,4^-),\,\cdots \sim z^4,\qquad\qquad\qquad\quad \text{for all BCFW pairs.}&\nonumber
\end{align}

Several remarks are given below:

1. The BCFW deformations of all non-vanishing four-point amplitudes $\mathcal{A}^{(4)}_6$ do not vanish in the large $z$ limit. As a result, the $S^{(3)}_6$ theory is not BCFW constructible \cite{Benincasa:2007xk}, and four-point amplitudes $\mathcal{A}^{(4)}_6$ cannot be constructed by gluing solely various copies of on-shell three-point functions $\mathcal{A}^{(3)}_6$. This is different from GR, which is, on the other hand, BCFW constructible \cite{Cachazo:2005ca,Benincasa:2007qj}. Generally, for BCFW inconstructible theories the on-shell recursion relations can be systematically written as
\begin{equation}
\mathcal{A}^{(n)}(0) = \sum\limits_{k}\frac{\mathcal{A}_L(z_k)\times\mathcal{A}_R(z_k)}{P^2_k(0)} + C_0,
\label{BCFWRecursion}
\end{equation}
where $k$ is some subset of the $n$ momenta. Aside from the standard on-shell recursive contributions from gluing various copies of lower-point amplitudes, we have to take into considerations the extra boundary contribution $C_0$ to obtain correct results.

2. It is, at the first sight, quite surprising that $\mathcal{A}^{(4)}_6(1^-,2^-,3^-,4^-)\neq0$ in the $S^{(3)}_6$ theory, as the naive analysis of the helicity structure of the three-point amplitudes $\mathcal{A}^{(3)}_6$ would show that there is no contribution from on-shell recursions. In other words, $\mathcal{A}^{(4)}_6(1^-,2^-,3^-,4^-)$ receives only the boundary contribution $C_0$, which is different from BCFW constructible theories like Yang-Mills theory and GR, where non-vanishing amplitudes receive contributions merely from on-shell recursions. Similarly, it is straightforward to show that, in Scherk-Schwarz theory the color-ordered 4-point amplitude $\mathcal{A}^{(4)}_{\text{SS}}(1^-,2^-,3^-,4^-)$ is also nonzero. It is interesting to study further the properties of scattering amplitudes that receive contributions merely from the boundary. 

\section{Non-Renormalization} 
\label{NonRenormalization}

In this section, we shall discuss the non-renormalization properties of higher derivative theories under investigation. Explicitly, we would like to show that quantum mechanical loops cannot correct the $S^{(3)}_4$ and $S^{(3)}_6$ vertices. This could be proved by using power counting, similar to proofs of non-renormalization theorems for Galileons, GR, $P(X)$ theories and conformal dilatons \cite{Goon:2016ihr}. 

For the $S^{(3)}_4$ theory in $D>4$, the momentum dependence of an $n$-point scattering amplitude $\mathcal{A}^{(n)}_4$ can be easily estimated as follows. Every loop integral leads to an integration $\sim \int\mathrm{d}^Dk$, every internal line contributes $\sim\frac{1}{k^2}$, and each copy of $S^{(3)}_4$ vertices contributes $\sim k^4$. Denoting the number of loops in the target Feynman diagram by $L$, the number of internal lines by $I$, and the number of copies of $S^{(3)}_4$ vertices by $V^{(3)}_4$, we find that the $n$-point amplitude $\mathcal{A}^{(n)}_4$ scales as $\sim k^{DL-2I+4V^{(3)}_4}$. This result can be simplified further using simple graph-theoretical identities,
\begin{equation}
n + 2I = 3 V^{(3)}_4, \qquad L = 1 + I -V^{(3)}_4,\nonumber
\end{equation}
and we have
\begin{equation}
\mathcal{A}^{(n)}_4 \sim k^{(D+2)L + 2n - 2}.
\end{equation}
In the above derivations, we have implicitly used dimensional regularization or some other mass-independent regularization schemes in loop calculations. This is quite important for the validity of our non-renormalization theorems. Also, the power counting results should be thought of as potentially containing logarithmic factors $\sim\log(k^2/\mu^2)$, with $\mu$ being the regularization scale. As a result, it is straightforward to show that the $S^{(3)}_4$ vertices cannot be renormalized by loop corrections, as loop corrections with $L\ge1$ renormalize instead multi-point vertices with derivatives $\ge D+6$.

Similarly, for the $S^{(3)}_6$ theory in $D\ge4$, we have
\begin{equation}
\mathcal{A}^{(n)}_6 \sim k^{(D+6)L + 4n -6},
\end{equation}
which means that loop corrections with $L\ge1$ renormalize multi-point vertices with derivatives $\ge D+12$. As a result, the $S^{(3)}_6$ vertices cannot be renormalized by loop corrections. 

Besides the $S^{(3)}_4$ and $S^{(3)}_6$ theory, it could be shown further that higher derivative theories with both $S^{(3)}_4$ and $S^{(3)}_6$ vertices also shares non-renormalization of the same type. For theories coupling to matter fields, we have found that in many cases there also exist similar non-renormalization theorems, thanks to the higher-derivative nature of various interacting vertices.

\section{Emergent Gravitons} 
\label{EmergentGravitons}

In this section, we would like to make a short discussion about the implications of higher derivative theories studied above on emergent gravitons from non-relativistic condensed matter systems. 

It has been known for some time that there could be emergent massless graviton excitations in the low-energy spectrum of various condensed matter systems, such as the $4D$ quantum hall system \cite{Zhang:2001xs} (see Ref.~\cite{Kraus:2013,Price:2015,Ozawa:2016htj} for experimental proposals), qubit models \cite{Gu:2009jh} and quantum nematic crystals \cite{Zaanen:2011hm}. A natural question one would like to ask at the next step could be \emph{what effective field theory describes interactions of these emergent gravitons}. The conventional guess is GR. However, we would like to stress that this is by no means a well-established physical fact, but theoretical prejudice based on so-called ``proofs'' of the uniqueness of GR as the only consistent theories for interacting massless gravitons. Based on the above studies of higher derivative theories of massless gravitons, we would like to propose these higher derivative theories instead as a possible answer. Similar to GR and many of its modifications, these higher derivative theories also respect Lorentz invariance and quantum mechanics. The violation of asymptotic causality resulted from the non-vanishing three-point vertices might not be a real problem for the case of emergent gravitons, as it could be traced back to the fact that the underlying condensed matter system does not respect relativistic causality from the very beginning.



\section{Summary} 
\label{Summary}

In this note, we have studied various properties of a novel class of higher derivative theories for interacting massless gravitons in Minkowski spacetime. Explicitly, we have studied their Lagrangian construction, violation of asymptotic causality, scattering amplitudes, non-renormalization properties, and implications on emergent gravitons in condensed matter systems. Besides open questions scattered in above discussions, we would like to end this note with the following remarks: 

First, it would be interesting to study possible supersymmetrizations of these higher derivative theories. It is known in literature \cite{Camanho:2014apa} that the three-point amplitude $\mathcal{A}^{(3)}_6$ is incompatible with supersymmetry. As a result, the $S^{(3)}_6$ theory could not be supersymmetrized. Also, $\mathcal{A}^{(3)}_4$ is incompatible with maximal supersymmetry, but it could be compatible with half maximal supersymmetry. The explicit form of supersymmetric $S^{(3)}_4$ theory has not been worked out yet.

Second, it would be interesting to understand in more details of scattering amplitudes in the $S^{(3)}_6$ theory. It is not clear to us yet whether there are closed general formula for at least some specific $n$-point amplitudes, or whether there is a recursive formulation to build higher-point amplitudes from lower-point amplitudes, which could be used for practical calculations. Moreover, it is shown in Ref.~\cite{Broedel:2012rc} that the $R^3$ modification to GR could be constructed by double copies of the $F^3$ modification of Yang-Mills theory (See Ref.~\cite{He:2016iqi} for a recent discussion on this issue from the viewpoint of Cachazo-He-Yuan formalism \cite{Cachazo:2013gna,Cachazo:2013hca}). Although closely related, the technical problem we attack is a bit different, and it is not clear whether our higher derivative theories share similar properties.

We shall come back to these issues in future publications.

\section*{Acknowledgements}


We would like to thank Zhao-Long Gu, Song He, and Yue Huang for helpful discussions. In this work, we have used the Mathematica packages xTras \cite{Nutma:2013zea}, FeynRules \cite{Alloul:2013bka}, FeynArts \cite{Hahn:2000kx}, FormCalc \cite{Gross:2014ola}, S@M \cite{Maitre:2007jq} for symbolic calculations.

\appendix
\section{Linearized Diffeomorphism Invariance of Pseudo-Linear Terms}
\label{PseudoLinear}

Here we give a formal demonstration of why the $(n+1)$-th order expansion of $n$-th Lovelock term in Minkowski spacetime is invariant up to total derivatives under linearized diffeomorphism. These terms are named by Hinterbichler \cite{Hinterbichler:2013eza} as ``pseudo-linear'' terms. The starting point is the full diffeomorphism invariance. 

Consider the expansion of the action around an arbitrary background $g_{\mu\nu}=\epsilon h_{\mu\nu}+ f_{\mu\nu}$,
\begin{align}
    S[\epsilon h+f]=\sum_{i=0}^{\infty}\epsilon^i S_f^{(i)}[h]. 
\end{align}
Every $S_f^{(i)}$ is a symmetric $i$-tic form $S_f^{(i)}[h,h,\dots,h]$ on the linear space of all $h$ configurations. Next, consider $\phi_{\epsilon X}$ as the 1-parameter diffeomorphism subgroup generated by a vector field $X^\mu$ and parametrized by $\epsilon$, such that $X$ and $h$ are of the same order. 
\begin{align}
    \phi^*_{\epsilon X}g &=\exp[\epsilon \lie_X](\epsilon h +f)=\sum_{i=0}^{\infty}\frac{\epsilon^i}{i!}(\lie_X)^i(\epsilon h +f)  \nonumber\\
    &=f + \sum_{i=1}^{\infty}\frac{\epsilon^{i}}{(i-1)!}(\lie_X)^{i-1} (h +\frac{1}{i}\lie_X f)  \nonumber\\
    &\equiv f + \sum_{i=1}^{\infty}\epsilon^{i}\tilde{h}^{(i)},
\end{align}
where $\lie_X$ denotes Lie derivative with respect to $X$. The first few orders of the expansion looks like
\begin{equation}
    \tilde{h}^{(1)} = h+\lie_X f,   \ \ \ \  \ \ \ \tilde{h}^{(2)} = \lie_X(h+\frac{1}{2}\lie_X f).
\end{equation}
Full diffeomorphism invariance demands that $S[\phi^*g]=S[g]$, which gives perturbatively,
\begin{align}
    \sum_{i=0}^{\infty}\epsilon^i S_f^{(i)}[h] = \sum_{i=0}^{\infty}\epsilon^i S_f^{(i)}\left[ \sum_{i=1}^{\infty}\epsilon^{i}\tilde{h}^{(i)} \right].
\end{align}
Suppose the first non-vanishing positive order of the expansion is $k$, then we have
\begin{subequations}
\begin{align}
    \epsilon^k &: S_f^{(k)}[h] = S_f^{(k)}[\tilde{h}^{(1)}],    \\
    \epsilon^{k+1} &: S_f^{(k+1)}[h] =  S_f^{(k+1)}[\tilde{h}^{(1)}] + k S_f^{(k)}[\tilde{h}^{(1)},\tilde{h}^{(1)},\dots,\tilde{h}^{(2)}].   \label{sndOrder}
\end{align}
\end{subequations}
Note that $\tilde{h}^{(1)}$ is the linearized diffeomorphism transformation that we are interested in. So the leading order of the action is automatically linearized diffeomorphism invariant, and so is the higher-order contribution apart from the addition of a term determined by the lower order in the expansion.

Let's carry out the above analysis for the Lovelock terms (wrapped in the spacetime integrals)
\begin{align}
    S_{n}=\int\mathrm{d}^Dx\sqrt{-g}\ R_{[\mu_1\nu_1}{}^{\mu_1\nu_1}R_{\mu_2\nu_2}{}^{\mu_2\nu_2}\cdots R_{\mu_n\nu_n]}{}^{\mu_n\nu_n}.
\end{align}
In Minkowski spacetime ($f=\eta$), we have $R_{\mu\nu}{}^{\rho\sigma}$ to be $O(h)$, so the leading order of $S_{n}$ is $n$, with the non-vanishing contribution given by
\begin{equation}
    S_{n|\eta}^{(n)}[h]=2^n\int\mathrm{d}^Dx\, \partial_{[\mu_1}\partial^{\mu_1}h_{\nu_1}^{\nu_1} \partial_{\mu_2}\partial^{\mu_2}h_{\nu_2}^{\nu_2}\cdots  \partial_{\mu_n}\partial^{\mu_n}h_{\nu_n]}^{\nu_n},
\end{equation}
which accidentally is an integral of the total derivative. To put more precisely, any expression of the form $S_{n|\eta}^{(n)}[h_1,h_2,\dots,h_n]$ is an integral of a total derivative. This makes the last term of \eqref{sndOrder} vanish, rendering the next-to-leading-order expansion $S_{n|\eta}^{(n+1)}[h]$ to be linearized diffeomorphism invariant. 




\begin{thebibliography}{99}

\bibitem{Bjerrum-Bohr:2014zsa} 
  N.~E.~J.~Bjerrum-Bohr, J.~F.~Donoghue, B.~R.~Holstein, L.~Plant¨¦ and P.~Vanhove,
  ``Bending of Light in Quantum Gravity,''
  Phys.\ Rev.\ Lett.\  {\bf 114}, 061301 (2015)
  [arXiv:1410.7590 [hep-th]].

\bibitem{Bjerrum-Bohr:2015vda} 
  N.~E.~J.~Bjerrum-Bohr, J.~F.~Donoghue, B.~K.~El-Menoufi, B.~R.~Holstein, L.~Plant¨¦ and P.~Vanhove,
  ``The Equivalence Principle in a Quantum World,''
  Int.\ J.\ Mod.\ Phys.\ D {\bf 24}, 1544013 (2015)
  [arXiv:1505.04974 [hep-th]].
  
\bibitem{Bjerrum-Bohr:2016hpa} 
  N.~E.~J.~Bjerrum-Bohr, J.~F.~Donoghue, B.~R.~Holstein, L.~Plante and P.~Vanhove,
  ``Light-like Scattering in Quantum Gravity,''
  JHEP {\bf 1611}, 117 (2016)
  [arXiv:1609.07477 [hep-th]].
  
\bibitem{Bai:2016ivl} 
  D.~Bai and Y.~Huang,
  ``More on the Bending of Light in Quantum Gravity,''
  Phys.\ Rev.\ D {\bf 95}, 064045 (2017)
  [arXiv:1612.07629 [hep-th]].
  
\bibitem{Bjerrum-Bohr:2017dxw} 
  N.~E.~J.~Bjerrum-Bohr, B.~R.~Holstein, J.~F.~Donoghue, L.~Plant¨¦ and P.~Vanhove,
  ``Illuminating Light Bending,''
  PoS CORFU {\bf 2016}, 077 (2017)
  [arXiv:1704.01624 [gr-qc]].

\bibitem{Wald:1986bj} 
  R.~M.~Wald,
  ``Spin-2 Fields and General Covariance,''
  Phys.\ Rev.\ D {\bf 33}, 3613 (1986).
  
\bibitem{Cutler:1986dv} 
  C.~Cutler and R.~M.~Wald,
  ``A New Type of Gauge Invariance for a Collection of Massless Spin-2 Fields. 1. Existence and Uniqueness,''
  Class.\ Quant.\ Grav.\  {\bf 4}, 1267 (1987).

\bibitem{Wald:1986dw} 
  R.~M.~Wald,
  ``A New Type of Gauge Invariance for a Collection of Massless Spin-2 Fields. 2. Geometrical Interpretation,''
  Class.\ Quant.\ Grav.\  {\bf 4}, 1279 (1987).

\bibitem{Heiderich:1988yq} 
  K.~Heiderich and W.~Unruh,
  ``Spin-2 Fields, General Covariance, and Conformal Invariance,''
  Phys.\ Rev.\ D {\bf 38}, 490 (1988).

\bibitem{Heiderich:1990bb} 
  K.~R.~Heiderich and W.~G.~Unruh,
  ``Nonlinear, noncovariant spin two theories,''
  Phys.\ Rev.\ D {\bf 42}, 2057 (1990).

\bibitem{Boulanger:2000rq} 
  N.~Boulanger, T.~Damour, L.~Gualtieri and M.~Henneaux,
  ``Inconsistency of interacting, multigraviton theories,''
  Nucl.\ Phys.\ B {\bf 597}, 127 (2001)
  [hep-th/0007220].

\bibitem{Horava:2010zj} 
  P.~Horava and C.~M.~Melby-Thompson,
  ``General Covariance in Quantum Gravity at a Lifshitz Point,''
  Phys.\ Rev.\ D {\bf 82}, 064027 (2010)
  [arXiv:1007.2410 [hep-th]].
  
\bibitem{deRham:2012ew} 
  C.~de Rham, G.~Gabadadze, L.~Heisenberg and D.~Pirtskhalava,
  ``Nonrenormalization and naturalness in a class of scalar-tensor theories,''
  Phys.\ Rev.\ D {\bf 87}, 085017 (2013)
  [arXiv:1212.4128].

\bibitem{Hinterbichler:2013eza} 
  K.~Hinterbichler,
  ``Ghost-Free Derivative Interactions for a Massive Graviton,''
  JHEP {\bf 1310}, 102 (2013)
  [arXiv:1305.7227 [hep-th]].
  
  \bibitem{Hertzberg:2016djj} 
 M.~P.~Hertzberg,
 ``Gravitation, Causality, and Quantum Consistency,''
  arXiv:1610.03065 [hep-th].

\bibitem{Hertzberg:2017abn} 
 M.~P.~Hertzberg and M.~Sandora,
 ``General Relativity from Causality,''
  JHEP {\bf 1709}, 119 (2017)
  [arXiv:1702.07720 [hep-th]].


\bibitem{Camanho:2014apa} 
  X.~O.~Camanho, J.~D.~Edelstein, J.~Maldacena and A.~Zhiboedov,
  ``Causality Constraints on Corrections to the Graviton Three-Point Coupling,''
  JHEP {\bf 1602}, 020 (2016)
  [arXiv:1407.5597 [hep-th]].
  
\bibitem{Metsaev:2005ar} 
  R.~R.~Metsaev,
  ``Cubic interaction vertices of massive and massless higher spin fields,''
  Nucl.\ Phys.\ B {\bf 759}, 147 (2006)
  [hep-th/0512342].
  
\bibitem{Metsaev:1986yb} 
  R.~R.~Metsaev and A.~A.~Tseytlin,
  ``Curvature Cubed Terms in String Theory Effective Actions,''
  Phys.\ Lett.\ B {\bf 185}, 52 (1987).
  
\bibitem{Bai:2017dwf} 
  D.~Bai and Y.~H.~Xing,
  ``On the uniqueness of ghost-free special gravity,''
  Commun.\ Theor.\ Phys.\  {\bf 68}, 329 (2017)
  [arXiv:1702.05756 [hep-th]].

  
\bibitem{Lovelock:1971yv} 
  D.~Lovelock,
  ``The Einstein tensor and its generalizations,''
  J.\ Math.\ Phys.\  {\bf 12}, 498 (1971).
  
\bibitem{Lovelock:1972vz} 
  D.~Lovelock,
  ``The four-dimensionality of space and the einstein tensor,''
  J.\ Math.\ Phys.\  {\bf 13}, 874 (1972).
  
\bibitem{BeltranJimenez:2018tfy} 
  J.~Beltran Jimenez, J.~A.~R.~Cembranos and J.~M.~Sanchez Velazquez,
  ``On scalar and vector fields coupled to the energy-momentum tensor,''
  arXiv:1803.05832 [hep-th].

  
\bibitem{Shapiro:1964uw} 
  I.~I.~Shapiro,
  ``Fourth Test of General Relativity,''
  Phys.\ Rev.\ Lett.\  {\bf 13}, 789 (1964).
  
\bibitem{Gao:2000ga} 
  S.~Gao and R.~M.~Wald,
  ``Theorems on gravitational time delay and related issues,''
  Class.\ Quant.\ Grav.\  {\bf 17}, 4999 (2000)
  [gr-qc/0007021].
  
\bibitem{Deser:1969wk} 
  S.~Deser,
  ``Self-interaction and gauge invariance,''
  Gen.\ Rel.\ Grav.\  {\bf 1}, 9 (1970)
  [gr-qc/0411023].
  
\bibitem{Deser:1987uk} 
  S.~Deser,
  ``Gravity From Self-interaction in a Curved Background,''
  Class.\ Quant.\ Grav.\  {\bf 4}, L99 (1987).

\bibitem{Padmanabhan:2004xk} 
  T.~Padmanabhan,
  ``From gravitons to gravity: Myths and reality,''
  Int.\ J.\ Mod.\ Phys.\ D {\bf 17}, 367 (2008)
  [gr-qc/0409089].
  
\bibitem{Deser:2009fq} 
  S.~Deser,
  ``Gravity from self-interaction redux,''
  Gen.\ Rel.\ Grav.\  {\bf 42}, 641 (2010)
  [arXiv:0910.2975 [gr-qc]].
  
\bibitem{Bai:2016pae} 
  D.~Bai,
  ``Softness, Polynomial Boundedness and Amplitudes' Positivity,''
  EPL {\bf 120}, 21001 (2017)
  [arXiv:1607.07301 [hep-th]].

  
\bibitem{Scherk:1974ca} 
  J.~Scherk and J.~H.~Schwarz,
  ``Dual Models for Nonhadrons,''
  Nucl.\ Phys.\ B {\bf 81}, 118 (1974).

\bibitem{Britto:2004ap} 
  R.~Britto, F.~Cachazo and B.~Feng,
  ``New recursion relations for tree amplitudes of gluons,''
  Nucl.\ Phys.\ B {\bf 715}, 499 (2005)
  [hep-th/0412308].

\bibitem{Britto:2005fq} 
  R.~Britto, F.~Cachazo, B.~Feng and E.~Witten,
  ``Direct proof of tree-level recursion relation in Yang-Mills theory,''
  Phys.\ Rev.\ Lett.\  {\bf 94}, 181602 (2005)
  [hep-th/0501052].
  
\bibitem{Benincasa:2007xk} 
  P.~Benincasa and F.~Cachazo,
  ``Consistency Conditions on the S-Matrix of Massless Particles,''
  arXiv:0705.4305 [hep-th].

\bibitem{Cachazo:2005ca} 
  F.~Cachazo and P.~Svrcek,
  ``Tree level recursion relations in general relativity,''
  hep-th/0502160.

\bibitem{Benincasa:2007qj} 
  P.~Benincasa, C.~Boucher-Veronneau and F.~Cachazo,
  ``Taming Tree Amplitudes In General Relativity,''
  JHEP {\bf 0711}, 057 (2007)
  [hep-th/0702032 [hep-th]].
  
\bibitem{Goon:2016ihr} 
  G.~Goon, K.~Hinterbichler, A.~Joyce and M.~Trodden,
  ``Aspects of Galileon Non-Renormalization,''
  arXiv:1606.02295 [hep-th].
  
%
%
%
%

\bibitem{Zhang:2001xs} 
  S.~C.~Zhang and J.~P.~Hu,
  ``A Four-dimensional generalization of the quantum Hall effect,''
  Science {\bf 294}, 823 (2001)
  [cond-mat/0110572].
  
  \bibitem{Kraus:2013}
Y.~E.~Kraus, Z.~Ringel, and O.~Zilberberg,
``Four-Dimensional Quantum Hall Effect in a Two-Dimensional Quasicrystal''
Phys.\ Rev.\ Lett.\ {\bf 111}, 226401 (2013)
[arXiv:1302.2647 [cond-mat.mes-hall]].

  
\bibitem{Price:2015}
  H.~M.~Price, O.~Zilberberg, T.~Ozawa, I.~Carusotto, and N.~Goldman, 
  ``Four-Dimensional Quantum Hall Effect with Ultracold Atoms,''
  Phys.\ Rev.\ Lett.\ {\bf 115}, 195303 (2015)
  [arXiv:1505.04387 [cond-mat.quant-gas]].
  
\bibitem{Ozawa:2016htj} 
  T.~Ozawa, H.~M.~Price, N.~Goldman, O.~Zilberberg and I.~Carusotto,
  ``Synthetic dimensions in integrated photonics: From optical isolation to four-dimensional quantum Hall physics,''
  Phys.\ Rev.\ A {\bf 93}, 043827 (2016).

\bibitem{Gu:2009jh} 
  Z.~C.~Gu and X.~G.~Wen,
  ``Emergence of helicity $\pm$ 2 modes (gravitons) from qubit models,''
  Nucl.\ Phys.\ B {\bf 863}, 90 (2012)
  [arXiv:0907.1203 [gr-qc]].
  
\bibitem{Zaanen:2011hm} 
  J.~Zaanen and A.~J.~Beekman,
  ``The Emergence of gauge invariance: The Stay-at-home gauge versus local-global duality,''
  Annals Phys.\  {\bf 327}, 1146 (2012)
  [arXiv:1108.2791 [cond-mat.str-el]].
  
%
%
    
\bibitem{Broedel:2012rc} 
  J.~Broedel and L.~J.~Dixon,
  ``Color-kinematics duality and double-copy construction for amplitudes from higher-dimension operators,''
  JHEP {\bf 1210}, 091 (2012)
  [arXiv:1208.0876 [hep-th]].
  
\bibitem{He:2016iqi} 
  S.~He and Y.~Zhang,
  ``New Formulas for Amplitudes from Higher-Dimensional Operators,''
  arXiv:1608.08448 [hep-th].
  
\bibitem{Cachazo:2013gna} 
  F.~Cachazo, S.~He and E.~Y.~Yuan,
  ``Scattering equations and Kawai-Lewellen-Tye orthogonality,''
  Phys.\ Rev.\ D {\bf 90}, 065001 (2014)
  [arXiv:1306.6575 [hep-th]].

\bibitem{Cachazo:2013hca} 
  F.~Cachazo, S.~He and E.~Y.~Yuan,
  ``Scattering of Massless Particles in Arbitrary Dimensions,''
  Phys.\ Rev.\ Lett.\  {\bf 113}, 171601 (2014)
  [arXiv:1307.2199 [hep-th]].
  
\bibitem{Nutma:2013zea} 
  T.~Nutma,
  ``xTras : A field-theory inspired xAct package for mathematica,''
  Comput.\ Phys.\ Commun.\  {\bf 185}, 1719 (2014)
  [arXiv:1308.3493 [cs.SC]].
  
\bibitem{Alloul:2013bka} 
  A.~Alloul, N.~D.~Christensen, C.~Degrande, C.~Duhr and B.~Fuks,
  ``FeynRules  2.0 - A complete toolbox for tree-level phenomenology,''
  Comput.\ Phys.\ Commun.\  {\bf 185}, 2250 (2014)
  [arXiv:1310.1921 [hep-ph]].
  
\bibitem{Hahn:2000kx} 
  T.~Hahn,
  ``Generating Feynman diagrams and amplitudes with FeynArts 3,''
  Comput.\ Phys.\ Commun.\  {\bf 140}, 418 (2001)
  [hep-ph/0012260].

\bibitem{Gross:2014ola} 
  C.~Gro$\beta$, T.~Hahn, S.~Heinemeyer, F.~von der Pahlen, H.~Rzehak and C.~Schappacher,
  ``New Developments in FormCalc 8.4,''
  PoS LL {\bf 2014}, 035 (2014)
  [arXiv:1407.0235 [hep-ph]].

\bibitem{Maitre:2007jq} 
  D.~Maitre and P.~Mastrolia,
  ``S@M, a Mathematica Implementation of the Spinor-Helicity Formalism,''
  Comput.\ Phys.\ Commun.\  {\bf 179}, 501 (2008)
  [arXiv:0710.5559 [hep-ph]].


\end{thebibliography}


\end{document}